\documentclass[twocolumn,
pra
]{revtex4}

\usepackage{graphicx}
\usepackage{amssymb}
\usepackage{amsmath}

\newcommand{\vect}[1]{{\mathbf #1}}

\def\lba{\left(}    \def\rba{\right)}
\def\lbc{\left[}    \def\rbc{\right]}

\begin{document}

\title{Magnetocapacitance without magnetism}

\author{Meera M. Parish}
\email{meera.parish@ucl.ac.uk} %
\affiliation{London Centre for Nanotechnology, Gordon Street, London,
  WC1H 0AH, United Kingdom}


\begin{abstract}
A substantial magnetodielectric effect is often an indication of
coupled magnetic and elastic order, such as is found in the
multiferroics.  However, it has recently been shown that magnetism is
not necessary to produce either a magnetoresistance or a
magnetocapacitance when the material is inhomogeneous [M. M. Parish
  and P. B. Littlewood, Phys.\ Rev.\ Lett.\ {\bf 101}, 166602
  (2008)]. 
Here, we will investigate the characteristic magnetic-field-dependent
dielectric response of such an inhomogeneous system using exact
calculations and numerical simulations of conductor-dielectric
composites.  
In particular, we will show that even simple conductor-dielectric
layers exhibit a magnetocapacitance, and thus random bulk
inhomogeneities are not a requirement for this effect. Indeed, this
work essentially provides a natural generalisation of the
Maxwell-Wagner effect to finite magnetic field. We will also discuss
how this phenomenon has already been observed experimentally in some
materials.  
\end{abstract}


\maketitle

\section{Introduction}

Macroscopic inhomogeneities can have a surprising impact on the
magnetotransport of a material. 
A spectacular example of this is the anomalous transverse 
magnetoresistance observed in doped silver
chalcogenides~\cite{silver1,silver2}. Here, the presence of 
inhomogeneities distorts the current paths in such a way that the Hall
part is mixed into the longitudinal part of the response, thus giving
rise to  a large, linear magnetoresistance that persists across a
range of magnetic fields and temperatures~\cite{parish2003,parish2005}.  
Since this is a classical effect that does not require magnetism, it 
is relatively insensitive to temperature and thus potentially useful for
technological applications. Indeed, inhomogeneities have already been
exploited in the design of magnetic sensors: semiconductor devices
with ``extraordinary magnetoresistance'' rely on the presence of a
metallic inclusion to distort current and increase the device's
resistance in a magnetic field~\cite{STHH00,ZHS01-2}. 

There are similar non-trivial effects 
in the case of the ac 
dielectric response: Catalan has shown that a material can display a
substantial magnetocapacitance 
when an intrinsic magnetoresistance is combined with  Maxwell-Wagner
extrinsic effects such as contact effects and bulk 
inhomogeneities~\cite{catalan2006}. Subsequently, Littlewood and I
showed that a magnetocapacitance can be induced by inhomogeneities
alone and magnetism is not even necessary~\cite{parish2008}.  
In particular, we found that a finite magnetic field induced a
characteristic dielectric \emph{resonance}, though no inductive
element was present in the system. 
As we discuss below, this is a counter-intuitive phenomenon that every 
experimentalist should be aware of when probing the dielectric
response of a material,  
particularly given that the existence of a
magnetocapacitance is often taken as an indication of coupled ferroelectric 
and magnetic order. 
It is already known that composites consisting of a ferroelectric and a
ferromagnet can generate a magnetoelectric coupling in the absence of any
intrinsic coupling~\cite{eerenstein2006}. 
Here, we examine how simple conductor-dielectric composites can yield a
dielectric response that depends on 
magnetic field in the absence of any magnetic or ferroelectric order.
This effect is insensitive to the microscopics and is thus generic,
being only dependent on the distribution of local capacitive and
conductive regions.  
The only relevant material parameters are the static dielectric
constant $\varepsilon$ of the capacitive regions, and the resistivity
tensor $\hat{\rho}$ (or conductivity tensor $\hat{\sigma} \equiv
\hat{\rho}^{-1}$) in the conducting regions.  
As such, 
this magnetodielectric effect can occur in a variety of inhomogeneous
materials:  
for instance, it has already been observed experimentally in 
nanoporous silicon~\cite{vasic2007,brooks2008} and potentially in 
the manganite La$_{2/3}$Ca$_{1/3}$MnO$_3$~\cite{rivas2006}, as
discussed in Sec.~\ref{sec:discussion}. 

In the following, we will investigate the characteristic
magnetic-field-dependent dielectric response 
of two-dimensional (2D) conductor-dielectric composites 
using exact calculations, numerical simulations, and the effective
medium approximation. In particular, we will show that even simple
conductor-dielectric layers exhibit a magnetocapacitance, 
and thus random inhomogeneities in the bulk are not a requirement for
this effect. Rather, it is sufficient to have conductor-dielectric
interfaces that are perpendicular to the flow of current (or the
movement of charges). This also appears to be consistent with
experiment, since there is an observed link between magnetocapacitance
and interfaces with free charges~\cite{maglione2008}.  
We will also discuss how an intrinsic magnetoresistance could enhance this effect.


\section{Model and methods}
\begin{figure}
\begin{center}
\includegraphics[width=0.9\linewidth,angle=0]{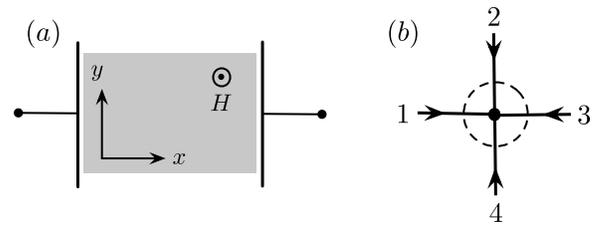}
\end{center}
\caption{The basics of probing and modelling dielectric response. 
Diagram (a) depicts the general measuring setup, where a 
rectangular sample is 
subjected to an oscillating electric field $E_x \hat{x}$ and 
a static transverse magnetic field $H$. The response of a general 2D
conductor-dielectric  
composite can be  
simulated with a network of four-terminal
elements (b).}
\label{fig:setup}
\end{figure}

We consider a classical 2D composite medium 
consisting of purely dielectric regions (defined by dielectric constant  
$\varepsilon$) and purely conducting regions. The latter regions have
the following resistivity tensor   
in a transverse magnetic field $\vect{H} = H\hat z$: 
\begin{align}\label{eq:rho}
\hat\rho =
 \left (
\begin{array}{cc}
\rho_{xx} & \rho_{xy}   \\
-\rho_{xy} & \rho_{xx} 
\end{array}
\right ) 
\end{align}
where we have assumed that the conductor is isotropic. We focus on the
simplest case where $\rho_{xx} = \rho$ and $\rho_{xy}/\rho = \mu H
\equiv \beta$, with $\mu$ the carrier mobility.  
Locally, the current $\vect{j}(\omega,\vect{r})$ is related to the
electric field $\vect{E}(\omega,\vect{r})$ via Ohm's law
$\vect{E}(\omega,\vect{r}) = \hat\rho(\omega,\vect{r})
\vect{j}(\omega,\vect{r}) \equiv
(i\omega\hat\varepsilon(\omega,\vect{r}))^{-1}
\vect{j}(\omega,\vect{r})$, and globally the system is driven by
electric field $\langle \vect{E}(\omega) \rangle$ at frequency
$\omega$, 
where $\langle ... \rangle$ corresponds to a volume average. 
The measured response averaged over the whole system is then the
effective resistivity defined from  
$\langle \vect{E}(\omega) \rangle = \hat\rho_e(\omega) \langle
\vect{j}(\omega) \rangle$. 
For the standard experimental setup where the boundary conditions are
such that $\langle j_y \rangle = 0$ as in Fig.~\ref{fig:setup}(a), the
components of the dielectric function that are actually probed  
are then given by $\varepsilon_{xx}(\omega) =
(i\omega\rho_{e,xx}(\omega))^{-1}$ and $\varepsilon_{xy}(\omega) =
(i\omega\rho_{e,xy}(\omega))^{-1}$. 
The longitudinal response $\varepsilon_{xx}(\omega)$ is 
the usual dielectric response measured in experiment, while the
transverse response $\varepsilon_{xy}(\omega)$ can be extracted from a
measurement of the transverse electric field $E_y(\omega) =
E_x(\omega) \varepsilon_{xx}(\omega)/\varepsilon_{xy}(\omega)$.  

For the isotropic two-component media considered in
Sec.~\ref{sec:media}, one can make use of the self-consistent
effective medium approximation outlined in
Refs.~\cite{S75,guttal2005,magier2006}. Here, we first imagine that we
have a single inclusion embedded in an effective medium and then we
average over all inclusions to self-consistently determine the
response of this effective medium. This amounts to solving the coupled
equations: 
\begin{align} \label{eq:SEMA}
\sum_i p_i (\hat{\sigma}_i - \hat{\sigma}_e) \lba 1 +
\frac{\hat{\sigma}_i - \hat{\sigma}_e}{2\sigma_{e,xx}} \rba = 0
\end{align}
where $\hat\sigma_e = \hat\rho_e^{-1}$ and $p_i$ is the volume
fraction of the $i$-th component. For the special case where the
fractions are equal ($p_1=p_2=1/2$), we can exploit a symmetry
transformation for the electric field and current
density~\cite{D71,B78,DR94} to derive an \emph{exact} result for the
effective dielectric function of the composite
medium~\cite{parish2008}. Note that the effective medium approximation
recovers this exact result for equal fractions, and becomes formally
exact in the limits $p_1 \to 0$, $p_2 \to 0$. Thus, we expect it to
provide an accurate approach for investigating isotropic media across
all volume fractions.  

For composites with anisotropic inhomogeneities like the
conductor-dielectric layers in Sec.~\ref{sec:layers}, one requires a
more brute force numerical approach. In this case, we discretise the
system into four-terminal elements as shown in
Fig.~\ref{fig:setup}(b), where the voltages $V_i$ at the terminals of
each element are linearly related to the currents $I_i$ via an
impedance matrix: $V_i = Z_{ij} I_j$. Note that each voltage is
defined with respect to the element centre and incoming currents are
defined as positive.  
In a purely conducting region, $Z_{ij} =
\frac{\rho}{2}(\delta_{ij}+\beta M_{ij})$, where  
\begin{align}
M_{ij}  = \frac{1}{2}\left(
\begin{array}{cccc}
0 & -1 & 0 & 1 \\
1 & 0 & -1 & 0 \\
0 & 1 & 0 & -1 \\
-1 & 0 & 1 & 0
\end{array}
\right) \ .
\end{align}
In a purely dielectric region, we simply have $Z_{ij} =
(2i\omega\varepsilon)^{-1} \delta_{ij}$.  
To simulate a given layed composite such as in Fig.~\ref{fig:layers}, 
we construct an $N\times M$ rectangular network of these four-terminal
elements and then take the limit of large $N$ and $M$ (keeping $N/M$
and the proportions of each phase fixed). In general, the response
converges rapidly to that of an infinite network.   

\section{Conductor-dielectric interfaces} \label{sec:layers}
\begin{figure}
\begin{center}
\includegraphics[width=0.9\linewidth,angle=0]{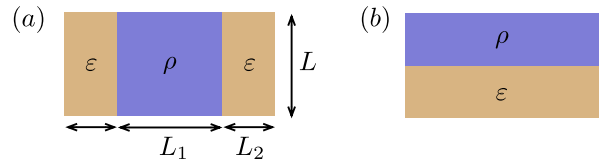}
\end{center}
\caption{Different configurations of conductor-dielectric layered
  composites, where  
$\langle j_y \rangle = 0$ on the upper and lower boundaries, and 
we have periodic boundary conditions for the left and right boundaries.
The interfaces between conductor $\rho$ and dielectric $\varepsilon$
can be perpendicular (a) or parallel (b) to the current 
flow.}
\label{fig:layers}
\end{figure}

In this section, we consider simple conductor-dielectric layers
(Fig.~\ref{fig:layers}), which constitute the simplest realisation of
anisotropic inhomogeneities. They also allow us to investigate the
behaviour of conductor-dielectric interfaces  
and they can thus 
potentially describe the effect of contacts in experiment. Indeed, a
magnetocapacitance has already been observed in non-magnetic Schottky
barriers~\cite{tongay2009}.

\begin{figure}
\begin{center}
\includegraphics[width=\linewidth,angle=0]{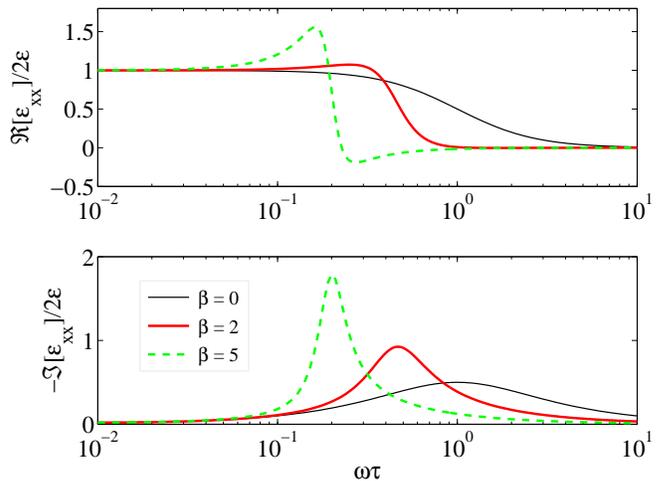}
\end{center}
\caption{Magnetodielectric effect in conductor-dielectric layers for
  the geometry in Fig.~\ref{fig:layers}(a) with $L_1 = 2L_2 =L/2$. 
For $\beta=0$ we have ordinary Maxwell-Wagner dielectric relaxation, while for $\beta > 1$ it becomes a dielectric resonance with characteristic frequency $\beta\omega\tau = 1$.} 
\label{fig:composites}
\end{figure}

In general, we find that the behaviour is strongly dependent on geometry.  
When the conductor-dielectric interfaces are parallel to the current
flow as in Fig.~\ref{fig:layers}(b), the dielectric response does not
depend on magnetic field and we simply obtain
$\varepsilon_{xx}(\omega) = \varepsilon + (i\omega\rho)^{-1}$. By
contrast, when the interfaces are perpendicular to the flow of charge
as in Fig.~\ref{fig:layers}(a), we have a substantial
magnetodielectric effect. Firstly, at zero magnetic field, we recover
the dielectric relaxation~\cite{jonscher1983} expected from the Maxwell-Wagner effect ---
for a square unit with equal proportions of each phase, i.e.\ $L_1 =
2L_2 =L/2$ in Fig.~\ref{fig:layers}(a), this gives 
\begin{align}
\varepsilon_{xx}(\omega) = 2\varepsilon \frac{1-i\omega\tau}{1+(\omega\tau)^2}
\end{align}
where the time constant $\tau = \rho\varepsilon$. Here, at the
characteristic frequency $\omega\tau = 1$, there is a rapid change in
$\Re[\varepsilon_{xx}]$ and an associated peak in
$\Im[\varepsilon_{xx}]$. Then, 
in the presence of a magnetic field, the peak shifts to smaller
$\omega\tau$ with increasing $\beta$ as shown in
Fig.~\ref{fig:composites}. Indeed, at large fields $\beta \gg 1$, the
characteristic frequency becomes $\beta\omega\tau = 1$ and the
behaviour evolves into a dielectric resonance, where the capacitance
$\Re[\varepsilon_{xx}]$ can become negative. This is similar to the
magnetocapacitance of isotropic composite media~\cite{parish2008} considered in Sec.~\ref{sec:media}.   
We also obtain a similar magnetodielectric effect for other
configurations of Fig.~\ref{fig:layers}(a), but the shift of the peak
position in $\Im[\varepsilon_{xx}]$ often only approximately obeys
$\beta\omega\tau = 1$. 

To gain insight into this effect, we consider a modified version of
the setup in Fig.~\ref{fig:layers}(a) where we insert a perfectly
conducting metallic interface between the conductor and
dielectric. Now, if we ignore the dielectric for the moment and
consider dc transport through the conductor, then such boundary
conditions will give rise to a magnetoresistance~\cite{pippard}. In
particular, if the conductor is square ($L_1=L$), 
then the effective resistivity is exactly $\rho^*(H) = \rho
\sqrt{1+\beta^2}$~\cite{IK92,parish2005}. 
Thus, the response of the conductor-dielectric composite becomes
(assuming $L_2=L/2$): 
\begin{align} \label{eq:interface}
\varepsilon_{xx}(\omega) = \frac{\varepsilon
  (1-i\omega\tau\sqrt{1+\beta^2})}{1+\omega^2\tau^2(1+\beta^2)} 
\end{align}
In the limit 
$\beta \gg1$, we essentially obtain Maxwell-Wagner relaxation with
$\omega\tau$ replaced  by 
$\beta\omega\tau$.    
Thus, we see that the effect of the interface is to mix the Hall component
into the response (both real and imaginary parts) and replace $\rho$ with
the Hall resistivity $\rho\beta$. Note that we do not obtain a dielectric resonance here and thus it appears that this feature requires non-zero Hall fields $E_y$  
within the dielectric itself, which is not the case for perfectly conducting interfaces.

\section{Isotropic composite media} \label{sec:media}
We now turn to the dielectric response of a 2D isotropic two-component medium where we once again have purely capacitive and purely resistive regions. 
When there are equal proportions of each component, we have the following exact results for the 
dielectric function~\cite{parish2008}
\begin{align}\label{eq:effdiel}
\varepsilon_{xx}(\omega) & = \varepsilon \frac{(1+i\omega\tau)}
{\sqrt{i\omega\tau}\sqrt{(1+i\omega\tau)^2-(\omega\tau\beta)^2}}
 \\ 
\varepsilon_{xy}(\omega) & =
\frac{\varepsilon}{\beta}\left(1-\frac{i}{\omega\tau}\right)
\end{align}
Note that $\varepsilon_{xx} \sim 1/\sqrt{\omega}$ in both low and high
frequency limits, in contrast to the layered case with equal proportions. 
However, for $\beta > 1$, we have a dielectric resonance phenomenon similar to Fig.~\ref{fig:composites}, with characteristic frequency $\beta\omega\tau = 1$.

We can also allow for the possibility that the resistive component has an intrinsic magnetoresistance $\rho_{xx}(H)$ by considering the more general expression for the resisitivity tensor in Eq.~\eqref{eq:rho}. In this case, we obtain
\begin{align} \label{eq:gendiel}
\varepsilon_{xx}(\omega) & =
\varepsilon\frac{1+i\omega\varepsilon\rho_{xx}}{\sqrt{i\omega\varepsilon\rho_{xx}}\sqrt{(1+i\omega\varepsilon\rho_{xx})^2 -\omega^2\varepsilon^2\rho_{xy}^2}} 
\end{align}
Here, we find that the size of  
the magnetocapacitance (i.e.\ the change of $\Re[\varepsilon_{xx}(\omega)]/\varepsilon$ with $\beta$) depends crucially on  
the ratio $\rho_{xy}/\rho_{xx}$. Thus, we can have an enhanced magnetocapacitance for a  \emph{negative} intrinsic magnetoresistance, where $\rho_{xy}/\rho_{xx} \sim \rho\beta/\rho_{xx}$ is large at finite magnetic field.
However, without a Hall component (i.e.\ $\rho_{xy} = 0$), we simply have
$\varepsilon_{xx}(\omega) =
\varepsilon/\sqrt{i\omega\epsilon\rho_{xx}}$ and we only have a
magnetocapacitance when there is an intrinsic magnetoresistance
$\rho_{xx}(H)$. In this case, we have no relaxation or resonance
phenomena. 

\subsection{Effective medium approximation}
To investigate all volume fractions of the components, we use the self-consistent effective medium approximation. 
Following Ref.~\cite{magier2006}, we simplify the problem by first
transforming to a frame where the resistivity tensors of each phase are
scalar. Thus we take
\begin{align}
\vect{E'} & = \vect{E} + b \hat{R} \rho \vect{J} \\ 
\vect{J'}  & = c\vect{J} + \hat{R} \rho^{-1} \vect{E}
\end{align}
where $\hat{R}$ is a rotation by 90 degrees:
\begin{align}
\hat{R} = \left( 
\begin{array}{cc}
0 & -1 \\
1 & 0 \end{array}
\right) \> ,
\end{align}
and the transformation constants are
\begin{align*}
b  = \ & \frac{1}{2\rho_{xy}\rho} \lbc \rho_{xx}^2 + \rho_{xy}^2 +
\frac{\rho^2}{(\omega \tau)^2}  \right. \\ 
& \left. + \sqrt{\lba \rho_{xx}^2 + \rho_{xy}^2 +
  \frac{\rho^2}{(\omega \tau)^2}  \rba^2 - 4
  \frac{\rho_{xy}^2\rho^2}{(\omega\tau)^2} } \rbc \\ 
c  = \ & - \frac{1}{2\rho_{xy}\rho} \lbc \rho_{xx}^2 + \rho_{xy}^2 +
\frac{\rho^2}{(\omega \tau)^2}  \right. \\ 
& \left. - \sqrt{\lba \rho_{xx}^2 + \rho_{xy}^2 +
  \frac{\rho^2}{(\omega \tau)^2}  \rba^2 - 4
  \frac{\rho_{xy}^2\rho^2}{(\omega\tau)^2} } \rbc \>, 
\end{align*}
using Eq.~\eqref{eq:rho} for the first component $\hat\rho_1$.
Note that these coefficients are real, even though the dielectric phase (component 2) has a purely
imaginary resistivity, $\rho_2 = (i\omega\varepsilon)^{-1}$. 
In the transformed frame, the (dimensionless) resistivities of the two
components are 
\begin{align*}
\frac{\rho'_1}{\rho} = \frac{\rho_{xx}}{\rho_{xy}+\rho c} \>, & &
\frac{\rho'_2}{\rho} = -\frac{i}{\omega\tau c}  
\end{align*}
From Eq.~\eqref{eq:SEMA}, 
the effective resistivity of the transformed composite is then 
\begin{align}\notag
\rho'_e = & \lba \frac{1}{2} - p\rba (\rho'_1 - \rho'_2) \\ 
& + \sqrt{\lba \frac{1}{2} - p\rba^2 (\rho'_1 - \rho'_2)^2 + \rho'_1 \rho_2'}
\end{align}
where $p$ is the volume fraction of the dielectric phase. 
Transforming back to the original frame then yields the final result
for the longitudinal effective resistivity: 
\begin{align} \label{eq:rhoSEMA}
\frac{\rho_{e,xx}}{\rho} = \frac{\rho\rho'_e}{\rho^2+(\rho'_e)^2} (b+c)
\end{align}
Note that taking equal proportions ($p=1/2$) yields the exact result
described earlier --- we can recover Eq.~\eqref{eq:gendiel} using
$\epsilon_{xx}(\omega) = (i\omega \rho_{e,xx})^{-1}$. 
In what follows, we assume that we simply have $\rho_{xx} = \rho$ and $\rho_{xy} = \beta\rho$.

For all volume fractions $p$, one can show that there is always a dielectric resonance at 
$\beta\omega\tau =1$ for large magnetic fields~\cite{parish2008}. Moreover, we find that the dissipation at the peak only depends on $p$ to leading order in $\beta$:
\begin{align}
\frac{\rho_{e,xx}}{\rho} \simeq \frac{2}{1-2p+\sqrt{1+4p(1-p)}} + iO(1/\beta)
\end{align}
where we see that $\rho_{e,xx}/\rho \to \infty$ when $p \to 1$, as expected.
Contrast this with the layered case (e.g.\ Eq.~\eqref{eq:interface}), where we find that $\rho_{e,xx}/\rho$ at the peak increases with increasing magnetic field $\beta$.

\begin{figure}
\begin{center}
\includegraphics[width=0.95\linewidth,angle=0]{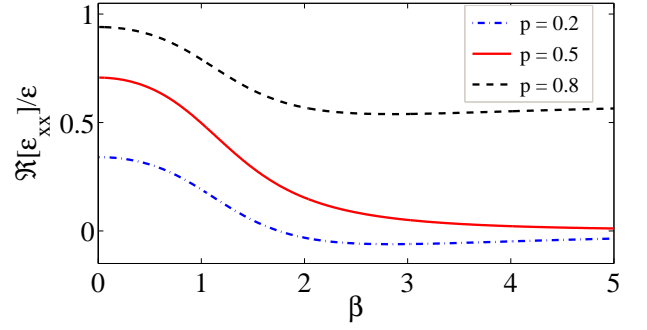}
\end{center}
\caption{Magnetocapacitance of isotropic conductor-dielectric composites at fixed frequency $\omega\tau=1$ for different dielectric volume fractions $p$. In the limit of high magnetic field $\beta \to \infty$, $\Re[\varepsilon_{xx}]/\varepsilon \to 0$ and $\Re[\varepsilon_{xx}]/\varepsilon \to 2p-1$ for $p \leq 1/2$ and $p>1/2$, respectively.}
\label{fig:MCiso}
\end{figure}

If we fix the frequency $\omega$, we also find a sizeable magnetocapacitance that depends on $p$. Figure~\ref{fig:MCiso} depicts the magnetocapacitance for $\omega\tau=1$, and
we see that both the size and sign of $\Re[\varepsilon_{xx}]/\varepsilon$ at large $\beta$ is determined by $p$. Indeed, in the limit $\beta \to \infty$, $\Re[\varepsilon_{xx}]/\varepsilon$ only remains finite when $p>1/2$ and the dielectric phase percolates through the whole sample.

Further insight can be obtained by deriving the limiting behaviour of the ac response from Eq.~\eqref{eq:rhoSEMA}. 
Lets first consider the case where the magnetic field $\beta = 0$. Then in the
limit $\omega\tau \to 0$, we have
\begin{align}\label{eq:lowfreq}
\frac{\rho_{e,xx}}{\rho} \simeq 
\begin{cases}
\frac{1}{1-2p} \ , &  p<1/2 \\ 
\frac{i}{\omega\tau} (1-2p) \ , &  p>1/2
\end{cases}
\end{align}
Likewise, in the limit $\omega\tau \to \infty$, we have
\begin{align}
\frac{\rho_{e,xx}}{\rho} \simeq 
\begin{cases}
1-2p \ , &  p<1/2 \\ 
\frac{i}{\omega\tau} \frac{1}{1-2p} \ , &  p>1/2
\end{cases}
\end{align}
Here, we can clearly see the effect of percolation, since the conducting regions dominate the transport below the percolation threshold of the dielectric ($p<1/2$), while the dielectric regions dominate above ($p>1/2$).
In particular, we see that the dependence on $p$ is inverted for low and high frequencies, since the dielectric regions effectively behave insulating in the former case and metallic in the latter (where there is insufficient time for the dielectric to build up charge).
However, for the special case $p=1/2$, i.e.\ at the percolation threshold,
we simply have $\rho_{e,xx} = \rho/\sqrt{i\omega\tau}$ and thus the
same behaviour in both limits.

When there is  
a large magnetic field $(\beta \gg 1)$, we obtain the
same result 
for $\omega\tau \to 0$, or
$\beta\omega\tau \ll 1$, but there can be markedly different behaviour when
$\beta\omega\tau \gtrsim 1$. Specifically, for $\beta \to \infty$ and finite $\omega\tau$,
we obtain
\begin{align}
\frac{\rho_{e,xx}}{\rho} \simeq 
\begin{cases}
\frac{1}{1-2p} \ , &  p<1/2 \\ 
\frac{\beta\sqrt{i\omega\tau}}{1+i\omega\tau} \ , & p=1/2 \\
\frac{i}{\omega\tau} \frac{1}{1-2p} \ , &  p>1/2
\end{cases}
\end{align}
which matches the large $\beta$ behaviour in Fig.~\ref{fig:MCiso}.
Note that for $p<1/2$, the system at high frequencies behaves like
one at low frequencies, Eq.~\eqref{eq:lowfreq}. This implies that the
dielectric region behaves insulating rather than metallic as one
might normally expect at high frequencies. This is because at large
magnetic field, the current paths are distorted such that they avoid any metallic region 
(cf.\ Ref.~\cite{STHH00})
and thus any metal effectively behaves like an insulator.

\section{Discussion} \label{sec:discussion}

\begin{figure}
\begin{center}
\includegraphics[width=0.95\linewidth,angle=0]{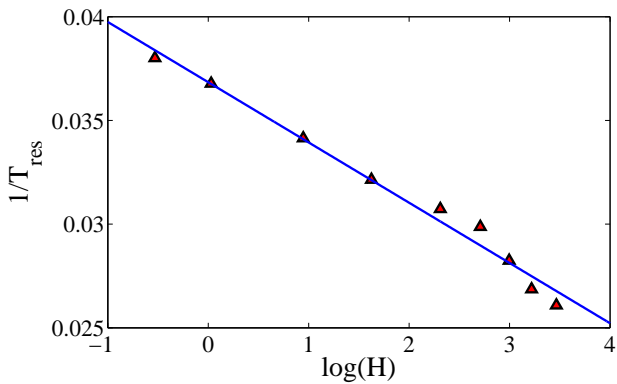}
\end{center}
\caption{Position of the resonance peak in the imaginary part of the
  dielectric response in terms of inverse temperature $1/T$ as a function of magnetic field $H$. Triangles are data for nanoporous silicon taken from Ref.~\cite{brooks2008}, while the line corresponds to a fit from our classical model 
of conductor-dielectric composites.
}
\label{fig:expt_fit}
\end{figure}

The 
dielectric response investigated in this paper only depends on the presence of macroscopic inhomogeneities and can thus potentially be observed in a variety of materials. There is already evidence to suggest that it accounts for the magnetodielectric effect in graphene-polyvinyl alcohol 
nanocomposite films~\cite{mitra2011} and in nickel nanosheet-Na-4 mica
composites~\cite{mitra2010}. 
Furthermore, our results are possibly relevant to rough interfaces in heterostructures~\cite{dussan2010}. 

Experiments on nanoporous silicon~\cite{vasic2007,brooks2008} have revealed a magnetic-field-dependent dielectric resonance that is consistent with our model. In this case, the dielectric response was measured at fixed frequency $\omega = 100$ kHz as a function of temperature $T=$ 25--40 K and magnetic field $H = $ 0--32 T.
However, 
varying the temperature is equivalent to varying $\omega\tau$ if we approximate $\varepsilon$ as temperature independent and the electrical transport in the semiconductor as being activated: $\rho = \rho_0 e^{\Delta/k_B T}$, where $\Delta$ is the activation gap. Then our predicted resonance occurs at the temperature $k_B T = -\Delta/\log(\beta\omega\varepsilon\rho_0)$, so that the dielectric resonance shifts to higher $T$ with increasing $H$, as indeed was observed in experiment. Assuming that $\beta = 1$ corresponds to $H\simeq$ 1T, we can fit the data with $\omega_0 \equiv 1/\rho_0\varepsilon \simeq 4 \times 10^{10}$ Hz and $\Delta \simeq 30$ meV, as shown in Fig.~\ref{fig:expt_fit}. 
Note that the size of $\Delta$ is consistent with activation from an impurity band.

Our effect may also be relevant to the dielectric resonance observed in the manganite La$_{2/3}$Ca$_{1/3}$MnO$_3$ just above the ferromagnetic transition temperature~\cite{rivas2006}. However, here the situation is more complex since there is magnetism involved and 
the effective ``composite'' corresponds to the phase separation between magnetic metal and charge-ordered insulator. Moreover, the size of magnetic metallic domains could change with magnetic field. This all requires further investigation: in principle, one might be able to probe the latter possibility by measuring the Hall component $\varepsilon_{xy}$, as discussed in Ref.~\cite{parish2008}. 

Finally, with regards to experiment, there is the question of when our classical model breaks down as the size of the sample is reduced. Formally, our approach is only valid when $1/\omega$ is greater than any microscopic timescale, e.g.,
the scattering time of charge carriers within the conducting regions, and when the size of the inhomogeneities is greater than any microscopic lengthscale, e.g., mean free path. However, one may be able to extend our results to smaller sizes using an appropriately redefined $\varepsilon$ and $\hat\rho$.

\section{Conclusion}
In this paper, we have investigated 2D conductor-dielectric composites and we have shown how these can produce a magnetic-field-dependent dielectric response without any magnetism. Using exact results and the effective medium approximation, we have derived expressions for the dielectric response of isotropic composite media and we have explicitly revealed the behaviour at the resonance $\beta\omega\tau =1$ and in the limits $\omega\tau \to 0$, $\omega\tau \to \infty$ and $\beta \to \infty$. Moreover, we find that an intrinsic magnetoresistance could enhance this magnetodielectric effect from inhomogeneities.
We have also performed numerical simulations of layered composites to show that conductor-dielectric interfaces play a key role in this phenomenon. 
Finally, we have discussed how our predicted magnetodielectric effect has already been observed experimentally in some materials.
An open question is how our results generalise to higher dimensions and this will be the subject of future work.

\acknowledgments
I am grateful to Peter Littlewood for fruitful discussions.  
This work was supported by the EPSRC under Grant No.\ EP/H00369X/2. 


\begin{thebibliography}{26}
\expandafter\ifx\csname natexlab\endcsname\relax\def\natexlab#1{#1}\fi
\expandafter\ifx\csname bibnamefont\endcsname\relax
  \def\bibnamefont#1{#1}\fi
\expandafter\ifx\csname bibfnamefont\endcsname\relax
  \def\bibfnamefont#1{#1}\fi
\expandafter\ifx\csname citenamefont\endcsname\relax
  \def\citenamefont#1{#1}\fi
\expandafter\ifx\csname url\endcsname\relax
  \def\url#1{\texttt{#1}}\fi
\expandafter\ifx\csname urlprefix\endcsname\relax\def\urlprefix{URL }\fi
\providecommand{\bibinfo}[2]{#2}
\providecommand{\eprint}[2][]{\url{#2}}

\bibitem[{\citenamefont{Xu et~al.}(1997)\citenamefont{Xu, Husmann, Rosenbaum,
  Saboungi, Enderby, and Littlewood}}]{silver1}
\bibinfo{author}{\bibfnamefont{R.}~\bibnamefont{Xu}},
  \bibinfo{author}{\bibfnamefont{A.}~\bibnamefont{Husmann}},
  \bibinfo{author}{\bibfnamefont{T.~F.} \bibnamefont{Rosenbaum}},
  \bibinfo{author}{\bibfnamefont{M.-L.} \bibnamefont{Saboungi}},
  \bibinfo{author}{\bibfnamefont{J.~E.} \bibnamefont{Enderby}},
  \bibnamefont{and} \bibinfo{author}{\bibfnamefont{P.~B.}
  \bibnamefont{Littlewood}}, \bibinfo{journal}{Nature}
  \textbf{\bibinfo{volume}{390}}, \bibinfo{pages}{57} (\bibinfo{year}{1997}).

\bibitem[{\citenamefont{Husmann et~al.}(2002)\citenamefont{Husmann, Betts,
  Boebinger, Migliori, Rosenbaum, and Saboungi}}]{silver2}
\bibinfo{author}{\bibfnamefont{A.}~\bibnamefont{Husmann}},
  \bibinfo{author}{\bibfnamefont{J.~B.} \bibnamefont{Betts}},
  \bibinfo{author}{\bibfnamefont{G.~S.} \bibnamefont{Boebinger}},
  \bibinfo{author}{\bibfnamefont{A.}~\bibnamefont{Migliori}},
  \bibinfo{author}{\bibfnamefont{T.~F.} \bibnamefont{Rosenbaum}},
  \bibnamefont{and} \bibinfo{author}{\bibfnamefont{M.-L.}
  \bibnamefont{Saboungi}}, \bibinfo{journal}{Nature}
  \textbf{\bibinfo{volume}{417}}, \bibinfo{pages}{421} (\bibinfo{year}{2002}).

\bibitem[{\citenamefont{Parish and Littlewood}(2003)}]{parish2003}
\bibinfo{author}{\bibfnamefont{M.~M.} \bibnamefont{Parish}} \bibnamefont{and}
  \bibinfo{author}{\bibfnamefont{P.~B.} \bibnamefont{Littlewood}},
  \bibinfo{journal}{Nature} \textbf{\bibinfo{volume}{426}},
  \bibinfo{pages}{162} (\bibinfo{year}{2003}).

\bibitem[{\citenamefont{Parish and Littlewood}(2005)}]{parish2005}
\bibinfo{author}{\bibfnamefont{M.~M.} \bibnamefont{Parish}} \bibnamefont{and}
  \bibinfo{author}{\bibfnamefont{P.~B.} \bibnamefont{Littlewood}},
  \bibinfo{journal}{Phys. Rev. B} \textbf{\bibinfo{volume}{72}},
  \bibinfo{pages}{094417} (\bibinfo{year}{2005}).

\bibitem[{\citenamefont{Solin et~al.}(2000)\citenamefont{Solin, Thio, Hines,
  and Heremans}}]{STHH00}
\bibinfo{author}{\bibfnamefont{S.~A.} \bibnamefont{Solin}},
  \bibinfo{author}{\bibfnamefont{T.}~\bibnamefont{Thio}},
  \bibinfo{author}{\bibfnamefont{D.~R.} \bibnamefont{Hines}}, \bibnamefont{and}
  \bibinfo{author}{\bibfnamefont{J.~J.} \bibnamefont{Heremans}},
  \bibinfo{journal}{Science} \textbf{\bibinfo{volume}{289}},
  \bibinfo{pages}{1530} (\bibinfo{year}{2000}).

\bibitem[{\citenamefont{Zhou et~al.}(2001)\citenamefont{Zhou, Solin, and
  Hines}}]{ZHS01-2}
\bibinfo{author}{\bibfnamefont{T.}~\bibnamefont{Zhou}},
  \bibinfo{author}{\bibfnamefont{S.~A.} \bibnamefont{Solin}}, \bibnamefont{and}
  \bibinfo{author}{\bibfnamefont{D.~R.} \bibnamefont{Hines}},
  \bibinfo{journal}{Appl.\ Phys.\ Lett.} \textbf{\bibinfo{volume}{78}},
  \bibinfo{pages}{667} (\bibinfo{year}{2001}).

\bibitem[{\citenamefont{Catalan}(2006)}]{catalan2006}
\bibinfo{author}{\bibfnamefont{G.}~\bibnamefont{Catalan}},
  \bibinfo{journal}{Appl. Phys. Lett.} \textbf{\bibinfo{volume}{88}},
  \bibinfo{pages}{102902} (\bibinfo{year}{2006}).

\bibitem[{\citenamefont{Parish and Littlewood}(2008)}]{parish2008}
\bibinfo{author}{\bibfnamefont{M.~M.} \bibnamefont{Parish}} \bibnamefont{and}
  \bibinfo{author}{\bibfnamefont{P.~B.} \bibnamefont{Littlewood}},
  \bibinfo{journal}{Phys. Rev. Lett.} \textbf{\bibinfo{volume}{101}},
  \bibinfo{pages}{166602} (\bibinfo{year}{2008}).

\bibitem[{\citenamefont{Eerenstein et~al.}(2006)\citenamefont{Eerenstein,
  Mathur, and Scott}}]{eerenstein2006}
\bibinfo{author}{\bibfnamefont{W.}~\bibnamefont{Eerenstein}},
  \bibinfo{author}{\bibfnamefont{N.~D.} \bibnamefont{Mathur}},
  \bibnamefont{and} \bibinfo{author}{\bibfnamefont{J.~F.} \bibnamefont{Scott}},
  \bibinfo{journal}{Nature} \textbf{\bibinfo{volume}{442}},
  \bibinfo{pages}{759} (\bibinfo{year}{2006}).

\bibitem[{\citenamefont{Vasic et~al.}(2007)\citenamefont{Vasic, Brooks,
  Jobiliong, Aravamudhan, Luongo, and Bhansali}}]{vasic2007}
\bibinfo{author}{\bibfnamefont{R.}~\bibnamefont{Vasic}},
  \bibinfo{author}{\bibfnamefont{J.~S.} \bibnamefont{Brooks}},
  \bibinfo{author}{\bibfnamefont{E.}~\bibnamefont{Jobiliong}},
  \bibinfo{author}{\bibfnamefont{S.}~\bibnamefont{Aravamudhan}},
  \bibinfo{author}{\bibfnamefont{K.}~\bibnamefont{Luongo}}, \bibnamefont{and}
  \bibinfo{author}{\bibfnamefont{S.}~\bibnamefont{Bhansali}},
  \bibinfo{journal}{Curr. Appl. Phys.} \textbf{\bibinfo{volume}{7}},
  \bibinfo{pages}{34} (\bibinfo{year}{2007}).

\bibitem[{\citenamefont{Brooks et~al.}(2008)\citenamefont{Brooks, Vasic,
  Kismarahardja, Steven, Tokumoto, Schlottmann, and Kelly}}]{brooks2008}
\bibinfo{author}{\bibfnamefont{J.~S.} \bibnamefont{Brooks}},
  \bibinfo{author}{\bibfnamefont{R.}~\bibnamefont{Vasic}},
  \bibinfo{author}{\bibfnamefont{A.}~\bibnamefont{Kismarahardja}},
  \bibinfo{author}{\bibfnamefont{E.}~\bibnamefont{Steven}},
  \bibinfo{author}{\bibfnamefont{T.}~\bibnamefont{Tokumoto}},
  \bibinfo{author}{\bibfnamefont{P.}~\bibnamefont{Schlottmann}},
  \bibnamefont{and} \bibinfo{author}{\bibfnamefont{S.}~\bibnamefont{Kelly}},
  \bibinfo{journal}{Phys. Rev. B} \textbf{\bibinfo{volume}{78}},
  \bibinfo{pages}{045205} (\bibinfo{year}{2008}).

\bibitem[{\citenamefont{Rivas et~al.}(2006)\citenamefont{Rivas, Mira,
  Rivas-Murias, Fondado, Dec, Kleemann, and Rodriguez}}]{rivas2006}
\bibinfo{author}{\bibfnamefont{J.}~\bibnamefont{Rivas}},
  \bibinfo{author}{\bibfnamefont{J.}~\bibnamefont{Mira}},
  \bibinfo{author}{\bibfnamefont{B.}~\bibnamefont{Rivas-Murias}},
  \bibinfo{author}{\bibfnamefont{A.}~\bibnamefont{Fondado}},
  \bibinfo{author}{\bibfnamefont{J.}~\bibnamefont{Dec}},
  \bibinfo{author}{\bibfnamefont{W.}~\bibnamefont{Kleemann}}, \bibnamefont{and}
  \bibinfo{author}{\bibfnamefont{M.~A.~S.} \bibnamefont{Rodriguez}},
  \bibinfo{journal}{Appl. Phys. Lett.} \textbf{\bibinfo{volume}{88}},
  \bibinfo{pages}{242906} (\bibinfo{year}{2006}).

\bibitem[{\citenamefont{Maglione}(2008)}]{maglione2008}
\bibinfo{author}{\bibfnamefont{M.}~\bibnamefont{Maglione}},
  \bibinfo{journal}{Journal of Physics: Condensed Matter}
  \textbf{\bibinfo{volume}{20}}, \bibinfo{pages}{322202}
  (\bibinfo{year}{2008}).

\bibitem[{\citenamefont{Stroud}(1975)}]{S75}
\bibinfo{author}{\bibfnamefont{D.}~\bibnamefont{Stroud}},
  \bibinfo{journal}{Phys.\ Rev.\ B} \textbf{\bibinfo{volume}{12}},
  \bibinfo{pages}{3368} (\bibinfo{year}{1975}).

\bibitem[{\citenamefont{Guttal and Stroud}(2005)}]{guttal2005}
\bibinfo{author}{\bibfnamefont{V.}~\bibnamefont{Guttal}} \bibnamefont{and}
  \bibinfo{author}{\bibfnamefont{D.}~\bibnamefont{Stroud}},
  \bibinfo{journal}{Phys. Rev. B} \textbf{\bibinfo{volume}{71}},
  \bibinfo{pages}{201304} (\bibinfo{year}{2005}).

\bibitem[{\citenamefont{Magier and Bergman}(2006)}]{magier2006}
\bibinfo{author}{\bibfnamefont{R.}~\bibnamefont{Magier}} \bibnamefont{and}
  \bibinfo{author}{\bibfnamefont{D.~J.} \bibnamefont{Bergman}},
  \bibinfo{journal}{Phys. Rev. B} \textbf{\bibinfo{volume}{74}},
  \bibinfo{eid}{094423} (\bibinfo{year}{2006}).

\bibitem[{\citenamefont{Dykhne}(1971)}]{D71}
\bibinfo{author}{\bibfnamefont{A.~M.} \bibnamefont{Dykhne}},
  \bibinfo{journal}{Sov.\ Phys.\ JETP} \textbf{\bibinfo{volume}{32}},
  \bibinfo{pages}{348} (\bibinfo{year}{1971}).

\bibitem[{\citenamefont{Balagurov}(1978)}]{B78}
\bibinfo{author}{\bibfnamefont{B.~Y.} \bibnamefont{Balagurov}},
  \bibinfo{journal}{Sov.\ Phys.\ Solid State} \textbf{\bibinfo{volume}{20}},
  \bibinfo{pages}{1922} (\bibinfo{year}{1978}).

\bibitem[{\citenamefont{Dykhne and Ruzin}(1994)}]{DR94}
\bibinfo{author}{\bibfnamefont{A.~M.} \bibnamefont{Dykhne}} \bibnamefont{and}
  \bibinfo{author}{\bibfnamefont{I.~M.} \bibnamefont{Ruzin}},
  \bibinfo{journal}{Phys.\ Rev.\ B} \textbf{\bibinfo{volume}{50}},
  \bibinfo{pages}{2369} (\bibinfo{year}{1994}).

\bibitem[{\citenamefont{Tongay et~al.}(2009)\citenamefont{Tongay, Hebard,
  Hikita, and Hwang}}]{tongay2009}
\bibinfo{author}{\bibfnamefont{S.}~\bibnamefont{Tongay}},
  \bibinfo{author}{\bibfnamefont{A.~F.} \bibnamefont{Hebard}},
  \bibinfo{author}{\bibfnamefont{Y.}~\bibnamefont{Hikita}}, \bibnamefont{and}
  \bibinfo{author}{\bibfnamefont{H.~Y.} \bibnamefont{Hwang}},
  \bibinfo{journal}{Phys. Rev. B} \textbf{\bibinfo{volume}{80}},
  \bibinfo{pages}{205324} (\bibinfo{year}{2009}).

\bibitem[{\citenamefont{Jonscher}(1983)}]{jonscher1983}
\bibinfo{author}{\bibfnamefont{A.~K.} \bibnamefont{Jonscher}},
  \emph{\bibinfo{title}{Dielectric relaxation in solids}}
  (\bibinfo{publisher}{Chelsea Dielectrics Press}, \bibinfo{address}{London},
  \bibinfo{year}{1983}).

\bibitem[{\citenamefont{Pippard}(1989)}]{pippard}
\bibinfo{author}{\bibfnamefont{A.~B.} \bibnamefont{Pippard}},
  \emph{\bibinfo{title}{Magnetoresistance in Metals}}
  (\bibinfo{publisher}{Cambridge University Press},
  \bibinfo{address}{Cambridge}, \bibinfo{year}{1989}).

\bibitem[{\citenamefont{Isichenko and Kalda}(1992)}]{IK92}
\bibinfo{author}{\bibfnamefont{M.~B.} \bibnamefont{Isichenko}}
  \bibnamefont{and} \bibinfo{author}{\bibfnamefont{J.}~\bibnamefont{Kalda}},
  \bibinfo{journal}{J. Moscow Phys. Soc.} \textbf{\bibinfo{volume}{2}},
  \bibinfo{pages}{55} (\bibinfo{year}{1992}).

\bibitem[{\citenamefont{Mitra et~al.}(2011)\citenamefont{Mitra, Mondal, Saha,
  Datta, Banerjee, and Chakravorty}}]{mitra2011}
\bibinfo{author}{\bibfnamefont{S.}~\bibnamefont{Mitra}},
  \bibinfo{author}{\bibfnamefont{O.}~\bibnamefont{Mondal}},
  \bibinfo{author}{\bibfnamefont{D.~R.} \bibnamefont{Saha}},
  \bibinfo{author}{\bibfnamefont{A.}~\bibnamefont{Datta}},
  \bibinfo{author}{\bibfnamefont{S.}~\bibnamefont{Banerjee}}, \bibnamefont{and}
  \bibinfo{author}{\bibfnamefont{D.}~\bibnamefont{Chakravorty}},
  \bibinfo{journal}{J. Phys. Chem. C} \textbf{\bibinfo{volume}{115}},
  \bibinfo{pages}{14285} (\bibinfo{year}{2011}).

\bibitem[{\citenamefont{Mitra et~al.}(2010)\citenamefont{Mitra, Mandal, Datta,
  Banerjee, and Chakravorty}}]{mitra2010}
\bibinfo{author}{\bibfnamefont{S.}~\bibnamefont{Mitra}},
  \bibinfo{author}{\bibfnamefont{A.}~\bibnamefont{Mandal}},
  \bibinfo{author}{\bibfnamefont{A.}~\bibnamefont{Datta}},
  \bibinfo{author}{\bibfnamefont{S.}~\bibnamefont{Banerjee}}, \bibnamefont{and}
  \bibinfo{author}{\bibfnamefont{D.}~\bibnamefont{Chakravorty}},
  \bibinfo{journal}{Europhys. Lett.} \textbf{\bibinfo{volume}{92}},
  \bibinfo{pages}{26003} (\bibinfo{year}{2010}).

\bibitem[{\citenamefont{Dussan et~al.}(2010)\citenamefont{Dussan, Kumar, Scott,
  and Katiyar}}]{dussan2010}
\bibinfo{author}{\bibfnamefont{S.}~\bibnamefont{Dussan}},
  \bibinfo{author}{\bibfnamefont{A.}~\bibnamefont{Kumar}},
  \bibinfo{author}{\bibfnamefont{J.~F.} \bibnamefont{Scott}}, \bibnamefont{and}
  \bibinfo{author}{\bibfnamefont{R.~S.} \bibnamefont{Katiyar}},
  \bibinfo{journal}{Appl. Phys. Lett.} \textbf{\bibinfo{volume}{96}},
  \bibinfo{eid}{072904}  (\bibinfo{year}{2010}).

\end{thebibliography}

\end{document}